\documentclass[doublecol,figures]{epl2} 

\pdfoutput=1
\usepackage{graphicx}
\usepackage{dcolumn}
\usepackage{bm}
\usepackage{color}
\usepackage{stmaryrd}
\usepackage{epstopdf}
\newcommand{\M}[1]{\ensuremath{\mathbf{#1}}}
\newcommand{\mgs}{\ensuremath{m_{\mathrm{gs}}}}

\title{Magnetic frustration in the context of pseudo-dipolar ionic disorder.}

\author{S. T. Banks\inst{1} \and S. T. Bramwell\inst{2, 3}}
\shortauthor{S. T. Banks \etal}

\institute{                    
  \inst{1} Department of Chemistry, University College London, 20, Gordon Street, London, WC1H 0AJ, United Kingdom.
  \inst{2} London Centre for Nanotechnology, University College London, 17 -- 19 Gordon Street, London WC1H 0AH, United Kingdom.
  \inst{3} Department of Physics and Astronomy, University College London -
  Gower Street, London, WC1E 6BT.
}

\abstract{
We consider an alternative to the usual spin glass paradigm for disordered
magnetism, consisting of the previously unstudied combination of frustrated
magnetic interactions and pseudo-dipolar disorder in spin positions.
We argue that this model represents a general limiting case for real
systems as well as a realistic
model for certain binary fluorides and oxides. Furthermore, it is of great
relevance to 
the highly topical subjects of the Coulomb phase and `charge ice'. We derive
an analytical solution for the ground state phase diagram of a model system
constructed in this paradigm and identify magnetic phases that remain
either disordered or partially ordered even at zero temperature. These
phases are of a hitherto unobserved type, but may be broadly classified as
either `spin liquids' or `semi-spin liquids' in contrast to the usual spin
glass or semi-spin glass. Numerical simulations  are used to show that the spin liquid phase
exhibits no spin glass transition at finite temperature, despite the combination of frustration 
and disorder. By mapping onto a model of
uncoupled loops of Ising spins, we show that the magnetic structure factor of
this phase acts, in the limit
$T\rightarrow0$, as a sensitive probe of the positional disorder
correlations. We suggest that this result can be generalized to more complex
systems, including experimental realizations of canonical spin glass models.
}

\begin{document}

\maketitle

\section{Introduction}
Disordered phases of matter may be categorized in terms of the relative
importance of frustrated interparticle interactions and quenched positional
disorder (`frustration plus disorder').  For example, if the frustrated
interactions are magnetic and the
positional disorder is uncorrelated then one arrives at a recipe for a spin
glass, a concept which underpins much thinking about the nature of disordered
states~\cite{fisher_and_hertz}. However, 
it seems reasonable to ask if the combination of frustration
with {\it correlated} positional disorder can lead to distinct behaviour that
is not encompassed in the usual spin glass paradigm.

Pseudo-dipolar positional disorder is characterised by a two-particle correlation function in reciprocal space,
$g({\bf q})$, that decays like a dipole-dipole interaction. It contrasts strongly with conventional disorder, 
in which $g({\bf q})$ decays like a screened Coulomb interaction (an exponentially
cut-off power law). Spontaneously generated pseudo-dipolar disorder occurs in ice rule
ferroelectrics~\cite{YoungbloodAxe} and a magnetic equivalent has 
been observed in spin ice~\cite{Fennell}. However, to
our knowledge, there has been no comprehensive theoretical study of the effect of
pseudo-dipolar positional disorder on magnetic properties (although Villain
has previously discussed the
topic in the context of insulating spin glasses~\cite{Villain}). In particular,
the effect of varying the relative strengths of competing exchange interactions within a
system possessing pseudo-dipolar positional disorder has remained an unsolved
problem. 
Such a study could be relevant, either ideally, or approximately, to
many real systems.
For example, a combination of ice-rules positional disorder
with magnetic frustration could be realised experimentally in certain inverse
spinel ferrites~\cite{anderson_1956_1008} or by the family of
fluoride pyrochlores, exemplified by CsNiCrF$_6$.
In these systems two ionic species (e.g. Ni$^{2+}$, Cr$^{3+}$) are distributed over the
pyrochlore lattice {and Anderson showed that by minimizing the Coulomb interaction
the distribution of ions should obey the ice rules (although the perfection of
the ice rules in these systems has not been experimentally determined).} However
it is not clear that spin glass states are a generic property of this
experimental class~\cite{Steiner,Alba}. Hence it is relevant to ask the questions, what kind of magnetic states should one ideally expect,
and what are their experimental signatures?  Here we consider the case of ideal ice-rules 
disorder. Such a study is not only of interest to illuminate an alternative to the
traditional spin glass paradigm but is also relevant to the concept of 
the Coulomb phase~\cite{henley_2010_179}, a general
consequence of pseudo-dipolar correlations. Furthermore, there is
currently much interest in `charge ice' \cite{Fulde,Udagawa,Ishizuka}, in which
mobile electrons can adopt ice-rule configurations, leading to fractional excitations~\cite{Fulde}.
Our system provides a counterpoint to the electron
system, in that has quenched or static charge disorder, which influences
dynamical spin degrees of freedom.

\section{The Model}
The model we study consists of equal numbers of
two species of classical Heisenberg spins $\M{s}_a$ and
$\M{s}_b$, randomly distributed across the pyrochlore 
lattice (a cubic array of corner-linked tetrahedra) and 
subject to the `ice rule' constraint that there are two
spins of each type on every tetrahedron. The spins are coupled by Heisenberg
exchange parameters that take one of 
three possible values, $J_a$, $J_b$ or $J_{ab}$ depending on whether 
the neighbouring spins are both of type $a$, both of type $b$ or one of each
type respectively.
In the case that all spin {lengths} and exchange parameters are identical, the model reduces to 
either a pyrochlore lattice ferromagnet or antiferromagnet. While the former
orders conventionally, the latter remains in a cooperative   
paramagnetic state down to $T=0$~\cite{moessner_1998_2929}
and gives rise to a distinctive pinch-point magnetic structure factor
indicative of pseudo-dipolar  spin-spin
correlations~\cite{{zinkin_1997_11786},
  canals_2001_1323,henley_2005_014424,isakov_2004_167204,Fennell}.

The case where spin and exchange parameters differ is
addressed by expressing the spin Hamiltonian as a sum over
contributions from individual tetrahedra,  
\begin{equation}
H = -\frac{1}{N_T}\sum_{\alpha=1}^{N_T}\left(\sum_{\langle i,j
\rangle_{\alpha}}J_{ij}^{\alpha}\M{s}_i^{\alpha}.\M{s}_j^{\alpha}\right).
\label{eq:Hplaq}
\end{equation}
Here $N_T$ is the number of tetrahedra and $\langle
i,j\rangle_{\alpha}$ indicates all pairs of spins $i$ 
and $j$ on plaquette (tetrahedron) $\alpha$ with exchange interaction
$J_{ij}^{\alpha}$.
\subsection{A single tetrahedron}
\begin{figure}
\onefigure[width=8cm]{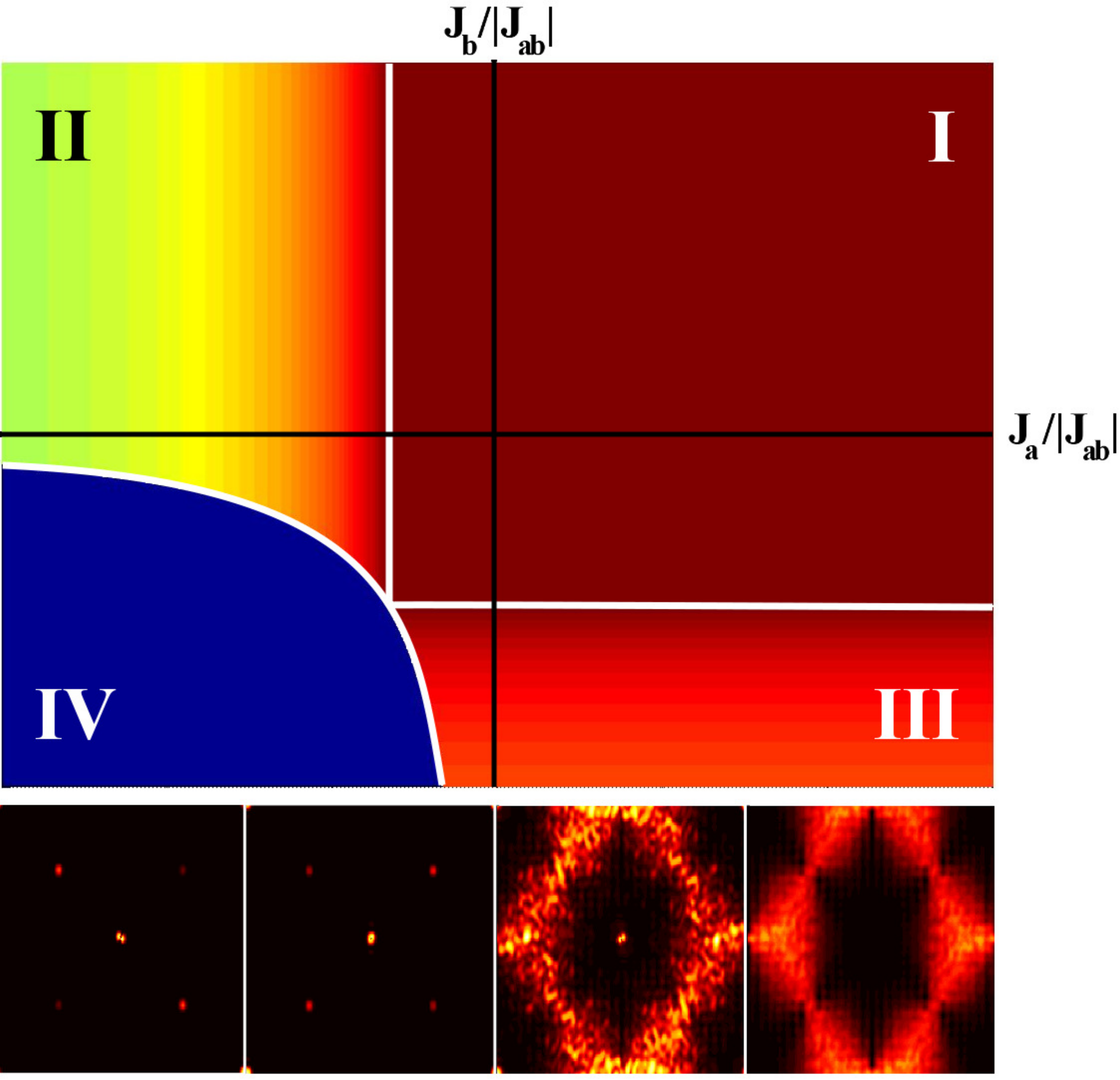}
\caption{\label{fig:phase_diag} The four regions of distinct magnetic
behaviour (boundaries in white) as a function of the reduced exchange
parameters $J_a/|J_{ab}|$ 
and $J_b/|J_{ab}|$. The boundary defining region IV is universal,
$|J_a||J_b|=|J_{ab}|^2$, while the
remaining boundaries are defined by the ratios of the spin lengths:
$J_a/|J_{ab}|=-s_b/s_a$ between regions I and II; $J_b/|J_{ab}|=-s_a/s_b$
between regions I and III. The background colour is an indication of the
{value of the}
ground state sublattice order parameter as defined by equation~\ref{eq:mgs1}.
The bottom panels are the magnetic neutron
scattering structure factors, $S(Q)$, in the different regions as discussed in
the text ((I-IV displayed left to right, respectively), shown
 in the $[h,h,l]$ reciprocal space plane from -2 to +2 on each
axis). For region II, scattering from all spins has been included 
whereas in region III only scattering from components of the $b$ spins
perpendicular to the common $a$ axis is shown {(the intensity in this third panel is
enhanced slightly relative to other regions in order to reveal the structure
of the diffuse scattering). The plots of $S(Q)$ have been obtained from Monte Carlo simulations
  at the points $(J_a,J_b,J_{ab})=(1,1,1)$, $(-2,2,1)$, $(2,-2,1)$ and
  $(-2,-2,1)$ for regions I, II, III and IV respectively (all at $T/|J_{ab}|=0.1$).}}
\end{figure}

We begin by focusing on a single tetrahedron and, with
no loss of generality, we assign spins $\M{s}_1$ and
$\M{s}_2$ to be of type $a$ and spins $\M{s}_3$ and $\M{s}_4$ to be of type $b$.
We then define three angular variables: $\phi_a$ as the angle between spins
$\M{s}_1$ and $\M{s}_2$; $\phi_b$ as the angle between spins $\M{s}_3$ and
$\M{s}_4$; $\theta$ as the angle 
between the resultants $\mathbf{S}_a=(\M{s}_1+\M{s}_2)$ and
$\mathbf{S}_b=(\M{s}_3+\M{s}_4)$. In these coordinates the Hamiltonian for the
single tetrahedron becomes
\begin{eqnarray}
H & = & -J_as_a^2\cos\phi_a-J_bs_b^2\cos\phi_b\nonumber\\
& &  -2J_{ab}s_as_b
(1+\cos\phi_a)^{\frac{1}{2}}(1+\cos\phi_b)^{\frac{1}{2}}\cos\theta.
\label{eq:ham_3}
\end{eqnarray}
The ground state magnetic configurations then separate into two
classes, 
characterized by the sign of $J_{ab}$, allowing an effective phase
diagram to be mapped out in the space spanned by the reduced variables
$J_a/|J_{ab}|$ and $J_b/|J_{ab}|$ (Fig.~\ref{fig:phase_diag}). The topology of
the resulting phase diagram is 
independent of the sign of $J_{ab}$, as are the equations of the
boundaries between regions.

In region I, spins of the same
species are aligned parallel. Spins of different species are either 
parallel or anti-parallel depending on the sign of $J_{ab}$. This behaviour extends even into 
the quadrants of $J$-space where one or both intra-species
interactions are antiferromagnetic. 
The extent of this domination by $J_{ab}$
is governed by the ratio $s_a/s_b$ as indicated by the equations of the
boundaries between region I and regions II and III (Figure~\ref{fig:phase_diag}).

Regions II and III are 
identical on interchanging the labels $a$ and $b$. 
In the ground state, spins of one species are perfectly parallel whilst the
antiferromagnetic coupling of the other species is frustrated by the
coupling $J_{ab}$. 
These spins cant away from the collinear axis defined by 
the first species through an angle {equal to half that between spins of type $x$,}
\begin{equation}
\cos\phi_x = \frac{2J_{ab}^2s_y^2}{|J_x|^2s_x^2}-1,
\label{eq:cant}
\end{equation}
where $\{x,y\}=\{a,b\}\,\,\, (\{b,a\})$ in region II (III) and $|J_x|>s_y/s_x$.
The single species sublattice order parameter is defined in these regions as
{
\begin{equation}
m^{(x)} =
\frac{1}{N_xs_x}\sqrt{\left(\sum_{i\in\{x\}}\M{s}_i\right).\left(\sum_{i\in\{x\}}\M{s}_i\right)}
\label{eq:mgs0}
\end{equation}
where $N_x$ is the number of spins of type $x$ and} the sum is over all spins of type $x$ (our 
definition anticipates extending the theory to a macroscopic
lattice). Combined with (\ref{eq:cant}) we see that 
the ground state order parameter is inversely proportional to $J_x$:
\begin{equation}
\mgs^{(x)} = \frac{s_y{|J_{ab}|}}{s_x|J_x|}.
\label{eq:mgs1}
\end{equation}
This order parameter is defined only in terms of the $x$ spins as the $y$
spins are perfectly ordered with respect to each other (parallel or
anti-parallel depending on the sign of $J_{ab}$) 
throughout regions II and III. Within regions I and IV we simply define
$\mgs^{(x)}=1,0$ respectively.

In region IV the intra-species
antiferromagnetic interactions dominate, leading to
configurations with $\phi_a=\phi_b=\pi$. Thus the third term of
(\ref{eq:ham_3}) is zero and the Hamiltonian is independent of $\theta$. In
this region, spins of different species are effectively decoupled. For
$J_{ab}>0$ the confluence of the phase boundaries is particularly interesting
as at this point the ordered ferromagnet becomes degenerate with the
antiferromagnetic spin liquid.

\subsection{Extension to the macroscopic lattice}
Consider now a macroscopic pyrochlore lattice having two A and two B ions per 
tetrahedron. A walker, starting at some ion of type A and following a path only
through sites populated by A ions (without retracing its steps) will always
return to its starting point. Furthermore, the path traced out will have no
branches but will form a continuous closed loop containing an even number of
lattice sites. The whole lattice is tiled with such loops -- every lattice site
belongs to one (and only one) of these closed, even membered, loops of spins of a single
species (referred to simply as `loops' from now on). The statistics of such loops have recently
been discussed in the context of the magnetic Coulomb phase~\cite{ludovic}. In
regions I, II and III, A type loops and B type loops interact with each other via
$J_{ab}$ which has the effect of imposing long range order. 
By contrast, in region IV there is no coupling between loops,
although spins within a given loop are perfectly
antiferromagnetically ordered with respect to each other. We consider this
soup of uncoupled, closed, antiferromagnetic loops to be a novel spin-liquid
like phase; frozen interactions within loops exist within a framework of two
mobile degrees of freedom per loop. There are no energy barriers to facilitate
global spin freezing and so, in the absence of free energy
barriers or dynamical constraints, there can be no spin glass transition in
region IV. This is in contrast to the Heisenberg pyrochlore
antiferromagnet (HPAFM) with weak random and uncorrelated bond
disorder~\cite{bellier_castella_2001_1365,saunders_2007_157201} to which a
number of points in region IV are closely related.
 
In all regions of $J$-space, the ground states of the single 
tetrahedron are robust to stacking, with no extra frustration
incurred. The phase diagram in Fig.~\ref{fig:phase_diag} should therefore be
equally valid for the macroscopic lattice. 

The decoupling of loops in region IV requires either
$\phi_a^{\alpha}=\pi$ or $\phi_b^{\alpha}=\pi$ on every tetrahedron
$\alpha$, both of which are true in the ground state. The individual loops 
may then be viewed as independent one-dimensional {Heisenberg}
chains, which may be arbitrarily long in the thermodynamic limit. For $T>0$,
$\phi_a^{\alpha}$ and $\phi_b^{\alpha}$ may differ from their ground state
values due to the excitation of low energy spin waves. 
The Hamiltonian then regains its dependence
on the variable $\theta$ which is likely to dramatically slow 
the dynamics. Our numerical simulations suggest, however, that this slowing
does not amount to truly broken ergodicity. Relaxing
the ice rules constraint on the ion placement prevents this dynamical slowing
down by providing extra unconstrained {magnetic} degrees of freedom
in plaquettes with all ions of the same species or ions in a 1:3
ratio. {It is readily shown that 1:3 and 3:1 tetrahedra place fewer constraints
on the magnetic degrees of freedom than do the 2:2 tetrahedra. This result
may be understood intuitively given that tetrahedra with all spins of the same species
are the least magnetically constrained of all.}

\section{Numerical Simulations}
Numerical evidence from Monte Carlo simulations is in agreement with our
theoretical predictions. Simulations were performed on lattices with
$L^3$ cubic unit cells ($L=4,7$, corresponding to 1024 spins and 5488 spins
respectively) for which a short loop algorithm~\cite{rahman_1972_4009} was 
used to generate ion 
configurations obeying the ice rules constraint. In all simulations we observed loops on all scales,
from the smallest possible (six membered
rings) up to loops spanning the system. With
one exception (discussed below) the energy scale was defined by $|J_{ab}|$ and
a single spin flip Metropolis algorithm was employed, with spin updates confined to a small
solid angle.

To investigate spin freezing, we recorded the
Edwards-Anderson order parameter
\begin{equation}
q_{\mathrm{EA}} =
\frac{1}{N}\sum_{i=1}^{N}\left\langle\mathbf{s}_i\right\rangle^2,
\end{equation}
in the region $0.01\le T/|J_{ab}|\le 0.1$ with $L=7$.
We chose {$s_a=3/2$, $s_b=1$}, (corresponding to the magnitudes of
the magnetic ions in CsNiCrF$_6$) and focused on the point
$(J_a/|J_{ab}|,J_b/|J_{ab}|)=(-1.1,-1.1)$, ensuring three different bond
contributions to the energy. We simulated both the completely random and ice
rules constrained models as described above. At each temperature the systems
were annealed in five steps from $T/|J_{ab}|=1$ with $10^6$ MCS/s (Monte Carlo
Steps per spin) ($10^7$ MCS/s were used at and below $T/|J_{ab}|=0.03$) for
equilibration at each step. Data was recorded over $10^6$ ($10^7$) MCS/s and averaged 
over ten disorder configurations. In the absence of the ice rules constraint
$q_{EA}$ is essentially zero at all the temperatures studied. Imposing
the constraint leads to a significant slowing of the dynamics, but with
$10^7$ MCS/s $q_{EA}$ remains below 0.05 even at $T/|J_{ab}|=0.01$.
As already noted, this behaviour is in strong contrast with that of HPAFM with
weak random bond
disorder~\cite{bellier_castella_2001_1365,saunders_2007_157201}.  

\begin{figure}
\onefigure[width=8cm]{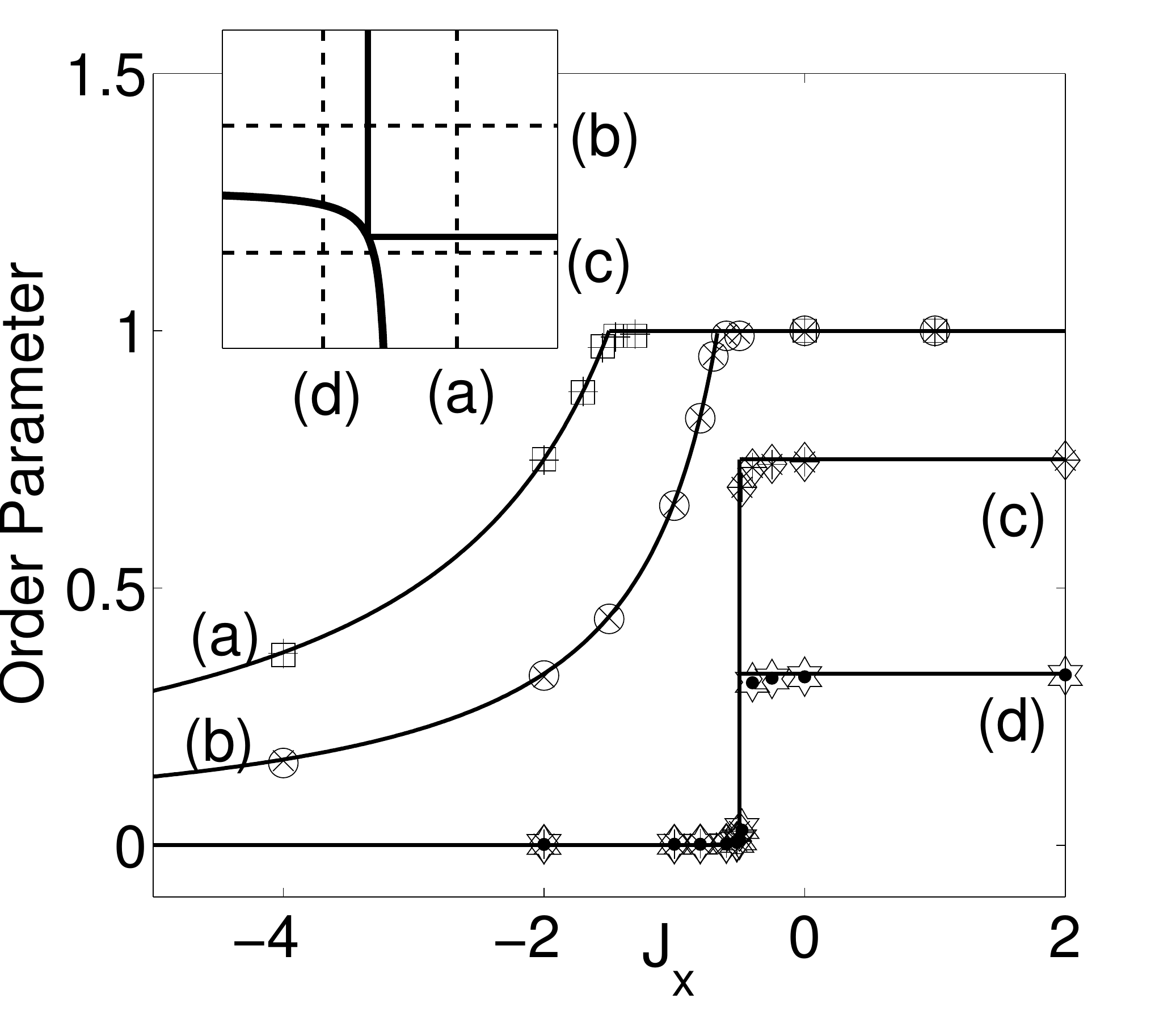}
\caption{
\label{fig:op}{Variations in the order parameter, $m^{(x)}$, along lines in
$J$-space denoted ($J_a/|J_{ab}|,J_b/|J_{ab}|$): (a) along
  $(2,J_b/|J_{ab}|)$; (b) along $(J_a/|J_{ab}|,2)$; (c) along
$(J_a/|J_{ab}|,-2)$; (d) along $(-2,J_b/|J_{ab}|)$. The solid lines represent the
  theoretical ground state magnetization as given by equation~\ref{eq:mgs1} in
  regions II and III (and defined as 1 in region I and 0 in region IV). The symbols represent the
  results of Monte Carlo simulations conducted at
  $T/|J_{ab}|=0.01$. Line symbols $+$, $\times$, $*$ and closed circles correspond to
  $J_{ab}>0$. Open symbols refer to $J_{ab}<0$.}}
\end{figure}

Figure~\ref{fig:op} shows the variation of the order parameter $m^{(x)}$ along the indicated
lines in $J$-space. The agreement between {the theoretical and
  numerical results} is striking
and this level of accuracy has been achieved with relatively small scale
simulations ($L=4$, {$s_a=3/2$, $s_b=1$} and MCS/s=$10^6$). These results
validate the extension of our  
analytical solution to the macroscopic lattice.

\begin{figure}
\onefigure[width=8cm]{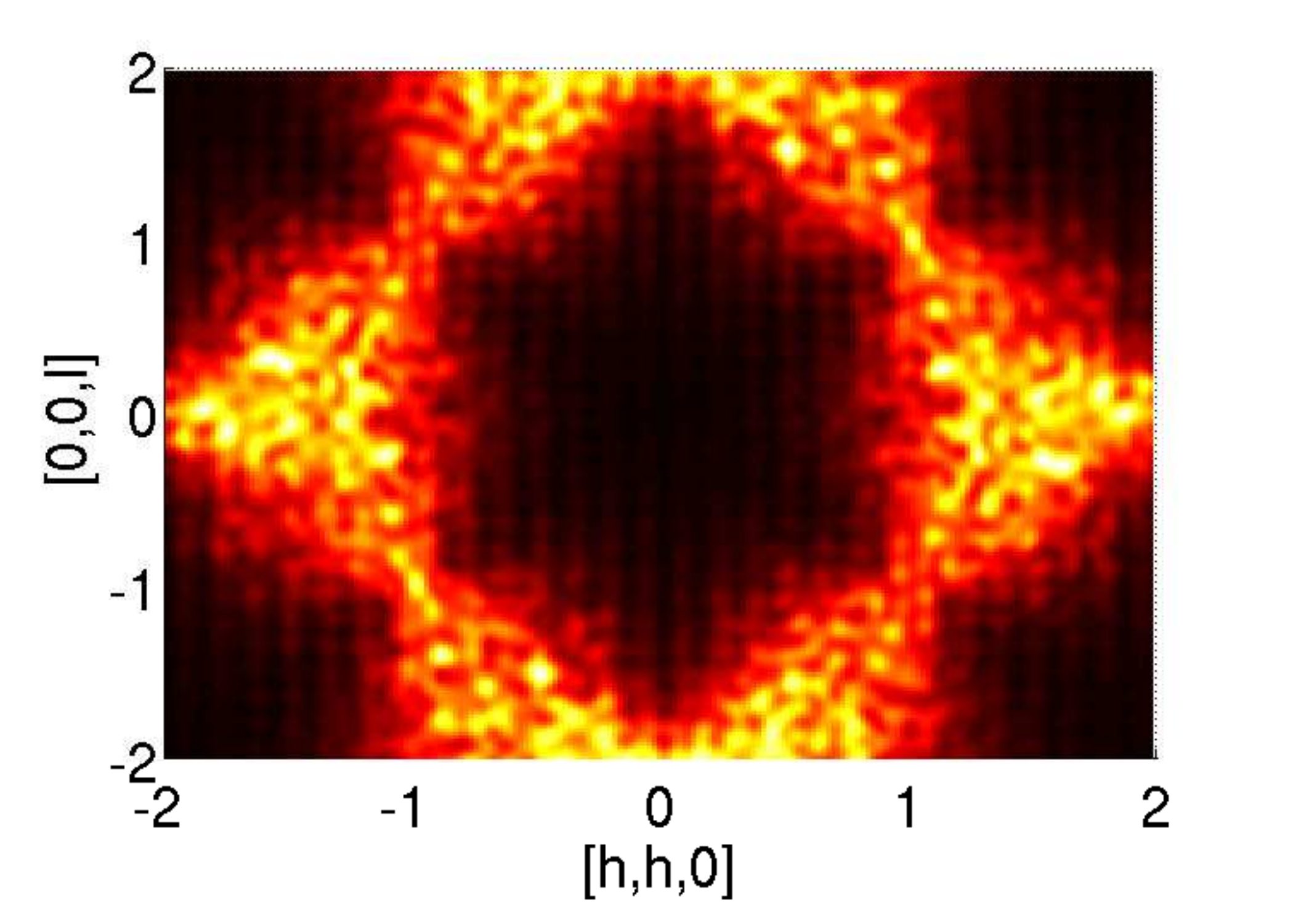}
\caption{
\label{fig:ideal-IV}{The magnetic structure factor $S(Q)$ for a system
tiled with independent chains of antiferromagnetically coupled Ising spins,
each chain with a randomly selected easy axis. This data was obtained using
$L=7$ with averaging over 10 easy axis configurations for each of 50 disorder
configurations.} }
\end{figure}

For the macroscopic lattice, differences in the ground state behaviour between
the regions should
manifest themselves in the magnetic structure factor, examples of which are
shown in the lower panels of Figure~\ref{fig:phase_diag}.  
For region I, long range
collinear order produces sharp Bragg peaks. In regions II and III,
sharp peaks arise from the 
components of all spins along the pseudo-collinear axis (shown for region II of
Figure~\ref{fig:phase_diag}), however scattering
from just the perpendicular component of the canted spins (shown for region III
of Figure~\ref{fig:phase_diag}) reveals
interesting diffuse scattering that suggests spin liquid like correlations
transverse to the ordered component. We suggest that this might be called a
`semi-spin liquid', in analogy with a semi-spin glass~\cite{Villain}.  
In region IV $S(\mathrm{Q})$ has the characteristic structure 
factor of an algebraic spin liquid, with pinch points indicative of the
pseudo-dipolar correlations, although without
the same clarity observed previously in studies of the HPAFM~\cite{canals_2001_1323,henley_2005_014424,isakov_2004_167204}. The
origins of such correlations in our model are not 
trivial. Unlike the pure HPAFM, the ground state in region IV has spins which
interact only within a single loop. For such a ground state to exhibit
pinch-point scattering would indicate that dipolar correlations emerge purely
as a consequence of the geometric distribution of loops, as governed by the
ion configuration. The spin-spin correlation 
function is then nothing more than the probability that the two spins
are on the same loop. To confirm this assertion we examined a toy model
representative of the ideal ground state of the 
system: $J_{ab}$ was set to zero and perfect N\'{e}el order was enforced
within each loop. We assigned a randomly selected easy axis to
each loop and measured $S(Q)$ for the resulting configuration. This 
process was repeated for a number of sets of randomly chosen easy axes and the
resulting structure factors averaged. The resolution was improved by
averaging again over a number of disorder configurations. The results
{(Figure~\ref{fig:ideal-IV})} clearly show signs of dipolar spin-spin correlations
emerging from this system of magnetically independent 1$d$ chains. To obtain
the data in this figure we averaged over 50 disorder configurations with
$L=7$. 
$S(Q)$ was averaged over 10 easy-axis configurations per
lattice. 


A direct consequence of the above observations is that the magnetic scattering
is acting as a probe of the structural disorder. 
There is however an inherent limit on the resolution of the pinch-point
scattering pattern which can result from a single realisation of the quenched
disorder on a finite lattice. Unlike the pure HPAFM, for which the bow-tie
pattern is well  resolved even for relatively small lattices, the ice rules
constrained binary pyrochlore described here has the lengths and spatial
arrangement of its loops predetermined by a particular
ion configuration.  The pure HPAFM however allows for a dynamic interchange of spins
between loops, in effect sampling a large number of loop configurations with
a corresponding increase in resolution. {In practical terms, this
  should have a similar effect to averaging the quenched disorder over many  
equivalent ion configurations. }

\section{Conclusions}
In conclusion, for the ideal model considered we have demonstrated the
suppression of spin glass behaviour and the emergence of novel spin liquid and semi-spin liquid 
phases. It will be of interest to re-examine the magnetic behaviour of the fluoride pyrochlores
in the light of this result. In particular, certain fluoride pyrochlores have been reported to show 
spin glass transitions~\cite{Steiner} however there is a growing body of
evidence that these compounds do not form traditional spin glasses below the
supposed freezing temperature~\cite{Alba, Ramirez_2, Harris}. We conclude that
if these are true spin glass transitions,  
they must be a consequence of disorder or interactions beyond those considered
here.
At a more general level we have illustrated a counter example to the idea that
geometric frustration and positional disorder must combine to generate a spin
glass, although we have not ruled out the possibility that  
the spin liquid states we have identified may be highly sensitive to further
quenched disorder of a different  
character. {Finally, our results have shed light on an issue of rather general importance
that is pertinent to
the interpretation of neutron scattering patterns of disordered magnets.
Thus, quenched atomic or ionic disorder is generally characterised by an energy
scale much higher than that of the magnetic interactions, so the magnetic
structure factor should generally be affected by the structural disorder correlations, but the question is to what degree, 
and what does this signify? We have identified a limiting case where the magnetic and structural disorder 
correlations are closely connected, and this connection is easily comprehended in terms of the model of `unfrustrated' spin loops we describe.
The opposite limiting case, which one would expect to apply to an ideal spin glass, is where the frustration is purely 
magnetic in origin and there is no dependence of magnetic on structural correlations. 
Real systems are likely to lie between these two limiting cases, so our results may be of some general relevance 
to the interpretation of neutron scattering patterns, including those of canonical spin glasses~\cite{murani_1999_131}
and weakly disordered spin glass systems such as  
`SCGO'~\cite{Ramirez} and Y$_2$Mo$_2$O$_7$~\cite{Gingras,Wiebe}. }




\acknowledgments
It is a pleasure to thank Mark Harris, Tom Fennell, Chris Henley, Peter
Holdsworth and John Chalker for very
stimulating discussions. S.\ T.\ Banks thanks the Ramsay Memorial Fellowship
Trust for funding through 
a Ramsay Memorial Fellowship.

\end{document}